\documentclass[twoside, 12pt, a4]{article}

\usepackage{titletoc}

\usepackage{graphicx, amsfonts, amsmath, amssymb, amscd,  theorem, exscale}

\usepackage{makeidx}
\makeindex

\newcounter{Figure}
\theoremstyle{plain}
\theoremheaderfont{\scshape}

\theorembodyfont{\upshape}

\theoremheaderfont{\scshape\small} \theorembodyfont{\scshape\small}

\newcommand{\be}{\begin{equation}}
\newcommand{\ee}{\end{equation}}
\newcommand{\bea}{\begin{eqnarray}}
\newcommand{\eea}{\end{eqnarray}}
\newcommand{\beas}{\begin{eqnarray*}}
\newcommand{\eeas}{\end{eqnarray*}}

\begin{document}
\begin{center}
{\Large \bf Who discovered the expanding universe?}
\end{center}
\vspace{5 pt}
\begin{center}
%\footnote{
%Harry Nussbaumer, ETH Zurich, Institute of Astronomy, Zurich, Switzerland. \\
%nussbaumer@astro.phys.ethz.ch \\
%Lydia Bieri, University of Michigan, Department of Mathematics, Ann Arbor MI, USA. \\
%lbieri@umich.edu
%}
{\large \bf  Harry Nussbaumer and Lydia Bieri   }
\end{center}

\vspace{14 pt}

Does it really matter who discovered the expanding universe? Great
discoveries are anyway never done single-handedly. This is a valid
attitude. However, those interested in the evolution of our
scientific culture are eager to know the intricate patterns that
lead to new insights. As the expanding universe is one of the most
important discoveries ever made, it is not astonishing that the
question of how it happened is still widely discussed.

The debate on this topic has flared up again due to an article in {\itshape Nature View}, where Eugenie Reich highlighted two contributions by Sidney van den Bergh \cite{vandenbergh} and David Block \cite{block}.
Their effect was to reanimate the discussion whether Hubble or Lema\^itre discovered the expanding universe, or whether it was simply a nearly predictable outcome of the normal scientific activity of those days. We have investigated this question in our book {\itshape Discovering the Expanding Universe} \cite{harrylydia}, where we reconstructed the discovery from original documents.

The story exemplifies two different paths of scientific progress. The Hubble-myth is that of a chance discovery, the story of Lema\^itre is a deliberate search by an individual scientist for the solution of a long standing problem. We draw attention to this aspect, and we also want to summarise the main facts for those too busy to do their own research, or to read our whole book \cite{harrylydia}, where also the detailed references can be found.

During the whole story we should not forget that neither Lema\^itre nor Hubble worked in isolation, and that a discovery also feeds on direct and indirect contributions from others. In this case the discovery was imbedded in the many contributions to general relativity, the search for an adequate interpretation of the enigmatic large nebular redshifts, and the challenge of measuring distances to extragalactic objects.

The facts are:
\begin{itemize}

\item[1.] In 1917 Einstein found from his fundamental equations of general relativity a static model of the universe.
More than two thousand years of astronomical observations showed the universe to be stable and practically immutable in space as well as in time.  Thus, very naturally, Einstein was looking for a static world.
To achieve his aim he introduced the cosmological constant $\Lambda$.

\item[2.] In 1917, a few months after Einstein, de Sitter found another solution which seemed to contain the explanation of Slipher's then already well known nebular redshifts, which he had been observing at Flagstaff since 1912. De Sitter's universe also contained the cosmological constant $\Lambda$, but no other matter.

\item[3.] In 1922 Friedmann found that Einstein's equations allowed a dynamic universe. He did not connect this finding to astronomical observations, and he did not spot the flaw in de Sitter's model. Except for Einstein, who did not think that dynamic solutions were physically relevant, no one took note of Friedmann.

\item[4.]There were other theoreticians, who in the 1920ies derived dynamical universes from the Einstein equations. But none of them linked their results to observations, nor did they propose an expanding universe. Lanczos in 1922 derived a formal solution of a spatially closed dynamical universe. In the same paper, he commented on publications by Hermann Weyl 1918 and 1919, which discussed redshifts in de Sitter's model. These papers attest to the confusion generated by de Sitter's empty universe. And as the mathematical tools had not yet been as developed as nowadays, interpretations often took intricate paths.
An important contribution came from Weyl with his concept of a ``causally connected world". It is also instructive to follow the Einstein-Weyl postcard exchange about the cosmological constant. More details about this topic and other players in the game can be found in our book \cite{harrylydia}.

\item[5.] During the early twenties not only the theoreticians but also the observers tried to make sense of de Sitter's universe, and to determine its radius of curvature. In addition, the community still debated whether nebulae were extragalactic or not. In the course of these investigations Carl Wirtz found in 1924 for spiral nebulae a relationship between their apparent photographic diameters and the radial velocities, and in the same year Knut Lundmark published in {\itshape Monthly Notices} the first distance-velocity diagram, distances being given in units of the distance to Andromeda. In the discussion on the nature of spiral nebulae \"Opik, by an ingenious method, had already in 1922 found a distance of 450 kpc to Andromeda, much closer to the real value than Hubble's later distance of 285 kpc. But it was Hubble's paper, read at the January 1, 1925 meeting of the American Astronomical Society, which cleared the sky for extragalactic nebulae (now called galaxies) as building blocks of the universe: the island universes hypothesised by Kant and Laplace were accepted as reality. For further details go to our book \cite{harrylydia}.

\item[6.] In 1927 Lema\^itre \cite{lem} critizised de Sitter's deadly sin for all the believers in the Copernican hypothesis: de Sitter's model violated the principle of homogeneity by treating the observer in a preferential way. Lema\^itre then discovered a set of dynamical solutions to Einstein's fundamental equations. From the theory of relativity he derived the linear velocity-distance relationship $v= H\cdot d$ (now called the Hubble relation).

Lema\^itre then connected his solution to observations. To derive the numerical value of $H$, Lema\^itre employed Hubble's 1926
distances to extragalactic nebulae and Slipher's redshifts. Depending on his choice of observations he arrived at either $625$ or $575  (km/s)/Mpc$ (compared to Hubble's  $500 (km/s)/Mpc$ in 1929). He was satisfied that the observations did not contradict his theoretical conclusions: the universe itself is expanding. But he was also well aware that there was an enormous scatter in the observations, and that further observations would have to confirm the linear relationship.

Lema\^itre was fully aware of the significance of his discovery. It is all the more astonishing that he did not try to place it in one of the prestigious astronomical journals, but published it in French in the {\itshape Annales de la Soci\'et\'e scientifique de Bruxelles}.

\item[7.] In 1929 Hubble \cite{hub} set out to study the motion of the sun against the background of the extragalactic nebulae, for which he tabulated, as others had done before, distances and velocities of extragalactic nebulae:
$v= rK + X \cos \alpha \cos \delta + Y \sin \alpha \cos \delta + Z \sin \delta$.
In the course of that investigation he found with his improved distances that ``The data in the table indicate a linear correlation between distances and velocities, whether the latter are used directly or corrected for solar motion, according to the older solutions". Having realised that fact, he turned away from the solar problem, concentrating on the linear relationship. Depending on how he grouped the galaxies, he found $K= 473, 513$ or $530 (km/s)/Mpc$, but opted for $K= 500 (km/s)/Mpc$ as his favourite value. To derive the numerical value of $K$, Hubble worked with his own distances. For the redshifts he mainly took those of Slipher, as tabulated in Eddington's 1923 {\itshape The Mathematical Theory of Relativity} (second edition 1924), without however, giving references. Hubble refrained from interpreting his observational discovery, he concluded ``The outstanding feature, however, is the possibility that the velocity-distance relation may represent the de Sitter effect, and hence that numerical data may be introduced into discussions of the general curvature of space". In a later letter to de Sitter, Hubble wrote that he would leave the interpretation of his observations to those ``competent to discuss the matter with authority". In none of the seven pages of Hubble's paper is there a single word about an expanding universe, actually Hubble never believed in such a thing. Hubble's observations confirmed Lema\^itre's predictions. In our book we also show how they re-ignited the cosmological debate, as exemplified by the crucial de Sitter-Eddington discussion of Friday, 10 January 1930 \cite{Observatory}.

\item[8.] Lema\^itre's article of 1927 appeared in French in the {\itshape Annales de la Soci\'et\'e scientifique de Bruxelles}. It was translated into English and published in 1931 in {\itshape Monthly Notices} \cite{lem1931}. However, there was a historically momentous omission. His derivation of the numerical value of $H$ was cut out by a deliberate act. 
Until recently it was an unsolved puzzle why this was done, and who was responsible. Thus the public, who read the English version, was left with the impression that Hubble had been the first to derive $H$. 
Hubble was even accused of having instigated the cuts in the translation. However, in 2011, two letters resolved the riddle. 

The first letter was a request to Lema\^itre by Dr. Smart on behalf of the Royal Astronomical 
Society (RAS) for a translation of his 1927 article for publication in the Monthly 
Notices (article \cite{block} by Block). The secretary of the RAS stresses that ``This request 
of the Council is almost unique in the Society's annals and it shows you how 
much the Society would appreciate the honour of giving your paper a 
greater publicity amongst English speaking scientists". Smart adds ``... if 
you have any further additions etc on the subject, we would glad[ly] print 
these too. I suppose that if there were additions a note could be inserted to 
the effect that \S\S1-72 are substantially from the Brussels paper + the 
remainder is new (or something more elegant)".        
Lema\^itre obliged (article \cite{liv2011} by Livio). In his answer to Smart of 13 February 1931 
he specified:``I did not find it advisable to reprint the provisional discussion of 
radial velocities which is clearly of no actual interest, ...". (For ``actual" he certainly 
had the French meaning of ``current" in mind.) 
He cut his derivation of the Hubble constant, which at that time was 
simply called the coefficient of expansion, and he cut the discussion of the 
astronomical data from which he had derived it. He replaced it with the
sentence: ``From a discussion of the available data, we adopt", after which he 
gives his numerical result for $\frac{R'}{R}$. However, he adds a reference list containing 
what for him must have been the crucial ``available data", namely the 1930 
series of observationally based papers by de Sitter. He certainly meant no 
offence to Hubble by not mentioning his 1929 publication, which in 1931 was 
outdated by de Sitter's thorough investigations. - Lema\^itre was a modest man. 
When his discoveries began to be attributed to Hubble, he refrained from a campaign to defend his priorities. 
However, in 1950, he reminded his readers 
that he had already determined the Hubble constant in 1927 \cite{lem1950}. 

\item[9.] It is occasionally stated that the mathematical prediction of the expanding universe was also made by Robertson. This is a misunderstanding which we also discuss in our book. In 1928 Robertson submitted to the {\itshape Philosophical Magazine} an article, in which he wanted to replace de Sitter's line element by ``a mathematically equivalent solution in which many of the apparent paradoxes inherent in [de Sitter's solution] were eliminated". He also arrived at the formula which in Lema\^itre's hand had become the distance-velocity relation. However, he wrote this as $v= c\cdot (l/R)$, where $l$ is the distance of
the nebula and $R$ the radius of curvature of the universe, for which he was looking within a static solution.

Robertson then took practically the same set of observations as had been taken by Lema\^itre one year before and would be taken by Hubble one year later. From this he calculated $R= 2 \cdot 10^{27}cm$. His $c/R$ corresponds to $H= 463 (km/s)/Mpc$; but this he did not calculate. Robertson placed an important milestone in our understanding of cosmological solutions of the Einstein equations. Solutions of Einstein's equations, in general, do not obey special symmetries. Yet, to describe the large scale structure of a spatially homogeneous universe,  the four-dimensional spacetime is usually separated into a spatial and a time component. Moreover, to `treat every point in this world equally' - the content of the Copernican principle -  and to implement the observational constraints into our models, leads to the hypothesis of universal homogeneity and isotropy.  These premises imply symmetries in the solutions of Einstein's equations. Robertson was the first to search in detail for all the mathematical universes that satisfy these physical requirements.

\item[10.] It is occasionally claimed that it was Hubble who converted Einstein to the expanding universe. This is very unlikely. Although there is no written report about the moment when Einstein was converted, it is highly probable that it happened, when Eddington showed him that his static solution was unstable. We discuss the circumstances further in our book.

\item[11.] Neither Hubble nor Lema\^itre rested on their laurels. With the help of the world's most powerful telescopes Hubble and Humason began measuring nebular redshifts on Mount Wilson. Their data - later continued by Sandage - would become one of the cornerstones of observational cosmology. Lema\^itre had another impact, when in 1931 \cite{lem1931c} he suggested in a one-column letter to Nature, what would become the Big Bang, and in 1933, in a paper read before the American National Academy of Sciences, Lema\^itre  suggested vacuum energy as the deeper meaning of the cosmological constant $\Lambda$. 
These exploits have also been highlighted in detail by Jean-Pierre Luminet \cite{luminet}. 

\item[12.] If Hubble was not the discoverer of the expanding universe, why is he still often venerated as such. Kragh and Smith
\cite{kraghsmith} have looked into the evolution of the `Hubble-myth'. They find that not until the 1950ies did the notion of `Hubble's law' and `Hubble as the astronomer who had discovered the expanding universe'  become common in the scientific literature, where Hubble's role was gradually elevated at the expense of everyone else's. They conclude: the label `Hubble's law' is an example of what has been called Stigler's law of eponymy, namely, `No scientific discovery is named after its original discoverer'.

\end{itemize}

The discovery of the expanding universe is a picture book example of an individual scientist who was aware of a burning scientific issue and solved it. It did not happen in a vacuum. Lema\^itre had benefitted from Eddington's insights into general relativity. In his 1927 paper he also cites Lanczos and Weyl, and he stood, of course, on the shoulders of Einstein and de Sitter. But similar arguments could be held against Newton and Einstein. However, if we apply our usual standards of attributing scientific discoveries, we should recall the situation of 1927. Einstein's static universe could not explain Slipher's redshifts, de Sitter's theory which provided redshifts was incomprehensible. Lema\^itre spotted the problem in de Sitter's work,
one of the great figures of  astronomy in the first half of the twentieth century. Before Lema\^itre only Friedmann had been sufficiently reckless to seriously follow up the idea of a truly dynamical universe. Now, in 1927, Lema\^itre derived from Einstein's fundamental equations the solution of a dynamical universe. To create a link to observations, he looked for the effect that his model would have on spectra of distant sources. This gave him the linear velocity-distance relationship $v=H \cdot d$, where a redshift signified an expanding universe, blueshifted spectra would have meant a shrinking universe. He then collected the available redshifts and distances to derive the missing factor of proportionality, which could not be derived from theory. The observations assured him that we live in an expanding universe. This was one of the most fascinating discoveries ever made.

The full story is much richer and more colourful than what can be
summarised on a few pages, and the following very incomplete list of
references is much extended in our book \cite{harrylydia}. \\ \\ \\ 

%\newpage 
%\noindent

\vspace{5 pt}

\noindent
Harry Nussbaumer \\ 
ETH Zurich \\ 
Institute for Astronomy \\ 
8093 Zurich, Switzerland \\
nussbaumer@astro.phys.ethz.ch \\ \\ 
Lydia Bieri \\ 
University of Michigan \\ 
Department of Mathematics \\ Ann Arbor MI 48109, USA. \\
lbieri@umich.edu


\begin{thebibliography}{99}
\bibitem{block} D. Block.
arxiv.org/abs/1106.3928 (2011).
\bibitem{hub} E. Hubble.
{\itshape A relation between distance and radial velocity among extra-galactic nebulae}, PNAS, 15, 168-173. (1929).
\bibitem{kraghsmith} H. Kragh and R.W. Smith.
{\itshape Who discovered the expanding universe?}, History of Science, Vol. 41, 141-162. (2003).
\bibitem{lem} G. Lema\^itre.
{\itshape Un univers homog\`ene de masse constante et de rayon croissant, rendant compte de la vitesse radiale des n\'ebuleuses extra-galctiques}, Annales de la Soci\'et\'e scientifique de Bruxelles, S\'erie A, 47, 49-59. (1927).
\bibitem{lem1931} G. Lema\^itre.
{\itshape A homogeneous universe of constant mass and increasing radius accounting for the radial velocity of extra-galactic nebulae},  MNRAS, 91, 483-490. (1931a).
\bibitem{lem1931c} G. Lema\^itre.
{\itshape The beginning of the world from the point of view of quantum theory},  Nature, 127, 706. (1931c). 
\bibitem{lem1950} G. Lema\^itre. 
{\itshape L'expansion de l'Univers, par Paul Couderc}. Annales
d'Astrophysique, 13, 344-345. (1950). 
\bibitem{liv2011} Livio, M. 
{\itshape Lost in translation: Mystery of the missing text solved.} 
 Nature 479, 171-173. (2011).
\bibitem{luminet} J.-P. Luminet.  
arxiv.org/abs/1105.6271v1 (2011). 
\bibitem{harrylydia} H. Nussbaumer and L. Bieri.
 {\itshape Discovering the Expanding Universe}, Cambridge University Press, Cambridge, UK. (2009).
ISBN 978-0-521-51484-2.
\bibitem{Observatory} {\itshape Proceeding of the R.A.S.}, The Observatory, Vol. 53, 37-39. (1930).
\bibitem{vandenbergh} S. van den Bergh.
arxiv.org/abs/1106.1195 (2011).
\end{thebibliography}
\end{document}